\documentclass{iau}
\usepackage{graphicx}

\title[Star-formation laws in extreme starbursts] %% give here short title %%
{Star-formation laws in extreme starbursts}
\author[S.~Garc\'{\i}a-Burillo, A.~Usero \& A.~Alonso-Herrero]   %% give here short author list %%
{S.~Garc\'{\i}a-Burillo$^1$, 
A.~Usero$^1$ \and 
A.~Alonso-Herrero$^2$}

\affiliation{$^1$ Observatorio de Madrid (OAN-IGN) \\ Alfonso XII, 3, 28014-Madrid, Spain \\ email: {\tt s.gburillo@oan.es, a.usero@oan.es} \\[\affilskip]
 $^2$ Instituto de F\'{\i}sica de Cantabria (IFCA-UC) \\ 39005-Santander, Spain \\ email: {\tt aalonso@ifca.unican.es}}

\pubyear{2012}
\volume{292}  %% insert here IAU Symposium No.
\pagerange{p1--p4}
\jname{Molecular gas, dust and star formation in galaxies}
\editors{Tony Wong \& Juergen Ott, eds.}
\begin{document}

\maketitle

\begin{abstract}

The observational study of star-formation laws is paramount to disentangling the physical processes at work on local and global scales in galaxies. To this aim we have expanded the sample of extreme starbursts, represented by local LIRGs and  ULIRGs, with high-quality data obtained in the 1-0 line of HCN.  The analysis of the new data shows that the star-formation efficiency of the dense molecular gas, derived from the FIR/HCN luminosity ratio,  is a factor 3-4 higher in extreme starbursts compared to normal galaxies. We find a duality in the Kennicutt-Schmidt  laws that is enhanced if we account for the different conversion factor for HCN ($\alpha_{\rm HCN}$) in extreme starbursts and correct for the unobscured star-formation rate in normal galaxies.   We find that it is possible to fit the observed differences in the FIR/HCN ratios between normal galaxies and LIRGs/ULIRGs with a common constant star-formation rate per free-fall time  (SFR$_{\rm ff}$) if we assume that HCN densities are $\sim$1--2 orders of magnitude  higher in LIRGs/ULIRGs, and provided  that SFR$_{\rm ff}$$\sim$0.005-0.01 and/or if $\alpha_{\rm HCN}$ is  a factor of a few lower than our favored values. 


\end{abstract}

\firstsection % if your document starts with a section,
\section{Star formation laws in galaxies: an observational perspective }

Based on simple  theoretical arguments, \cite[Schmidt~(1959)]{Sch59} proposed that the star-formation rate volume density ($\rho_{\rm SFR}$) in galaxies should scale as a power law  of the corresponding gas volume density ($\rho_{\rm gas}$): $\rho_{\rm SFR} \propto \rho_{\rm gas}^{n}$. If the relevant time scale for star formation is the 
{\it local} free-fall time (t$_{\rm ff}$), then $\rho_{\rm SFR} \propto \rho_{\rm gas}$/$t_{\rm ff}\propto \rho_{\rm gas}^{1.5}$.  A similar relation between the surface densities ($\Sigma_{\rm SFR}$, $\Sigma_{\rm gas}$) follows provided that the gas scale-height remains constant in different galaxy populations, a widely adopted yet far from trivial assumption.

Observers have used different proxies for  $\Sigma_{\rm SFR}$ and $\Sigma_{\rm gas}$ to derive star-formation relations in 
normal galaxies and extreme starbursts. \cite[Kennicutt et al.~(1998)]{Ken98} found a power law index $n$=1.40$\pm$0.15 using a compilation of 
galaxy-averaged CO+HI data obtained in a sample of $\sim$100 galaxies. More recently the CO surveys of \cite[Daddi et al.~(2010)]{Dad10} and \cite[Genzel et al.~(2010)]{Gen10} presented evidence of bimodality in galaxy-averaged star-formation laws: normal galaxies show 4--10 longer 
depletion time-scales compared to mergers. Observations of HCN lines have also been used to study the star-formation laws for the 
dense molecular gas.    The first studies by \cite[Gao \& Solomon~(2004)]{Gao04} found that the star-formation efficiency measured 
with respect to the dense molecular gas content (SFE$_{\rm{dense}} \propto L_{\rm{IR}}/L'_{\rm{HCN(1-0)}}$) was about constant in galaxies. This scenario was later questioned by \cite[Graci\'a-Carpio et al.~(2008)]{gc08}, who observed a sample of  LIRGs/ULIRGs 
and found that SFE$_{\rm dense}$ is a factor $\sim$2--3 higher in IR luminous targets compared  to normal galaxies. These 
observations suggested that galaxies classified as LIRGs may represent  the transition point in the star-formation laws.  To 
improve the statisitics in this luminosity range, we have recently expanded up to 28 the number of LIRGs with high-quality HCN data obtained with the IRAM 30m telescope. Out of these,  fourteen were extracted from the sample of nearby 
LIRGs studied by \cite[Alonso-Herrero et al.~(2006)]{Alo06}.  The results of this work, published by
\cite[Garc\'{\i}a-Burillo et al.~(2012)]{gb12}, have been used to study if the {\it observed} bimodality of star-formation laws in galaxies derived from 
CO  can be extended to the  higher density regime probed by HCN. We have also compared our observations with the predictions of  theoretical models in which the efficiency of star formation is determined by the ratio of a \emph{constant} star-formation rate per free-fall time (SFR$_{\rm ff}$) to the {\em local} free-fall time t$_{\rm ff}$ (\cite[Krumholz \& McKee~2005]{Kru05}).

\begin{figure*}[th!]
   \centering
   \includegraphics[width=6cm, angle=-90]{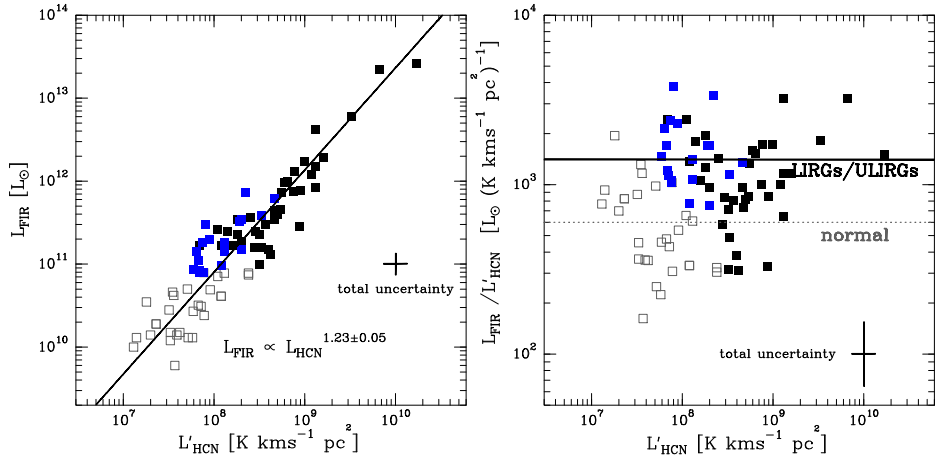}
   \caption{{\bf a)}~({\it left panel}) $L_{\rm FIR}$-$L'_{\rm HCN(1-0)}$ correlation plot derived for different galaxy samples. The solid line represents the 
   orthogonal regression fit to the full sample. {\bf b)}~({\it right panel}) The $L_{\rm FIR}/L'_{\rm HCN(1-0)}$ luminosity 
   ratio as a function of $L'_{\rm HCN}$. We display with different symbols normal galaxies  ($L_{\rm IR} < 10^{11} L_{\odot}$, open squares) and luminous infrared galaxies  
   ($L_{\rm IR} > 10^{11} L_{\odot}$, filled squares). The data from the new sample of LIRGs is identified by blue color markers. The horizontal lines  indicate  the 
   average value of $L_{\rm FIR}/L'_{\rm HCN(1-0)}$ in normal galaxies ($\sim$600$\pm$70$L_{\odot}\,$/(Kkms$^{-1}$pc$^2$)) and LIRGs/ULIRGs 
   ($\sim$1400$\pm$100$L_{\odot}\,$/(Kkms$^{-1}$pc$^2$)).} 
              \label{LFIR-LHCN}
\end{figure*}

\begin{figure*}[bth!]
   \centering
         \includegraphics[width=6cm, angle=-90]{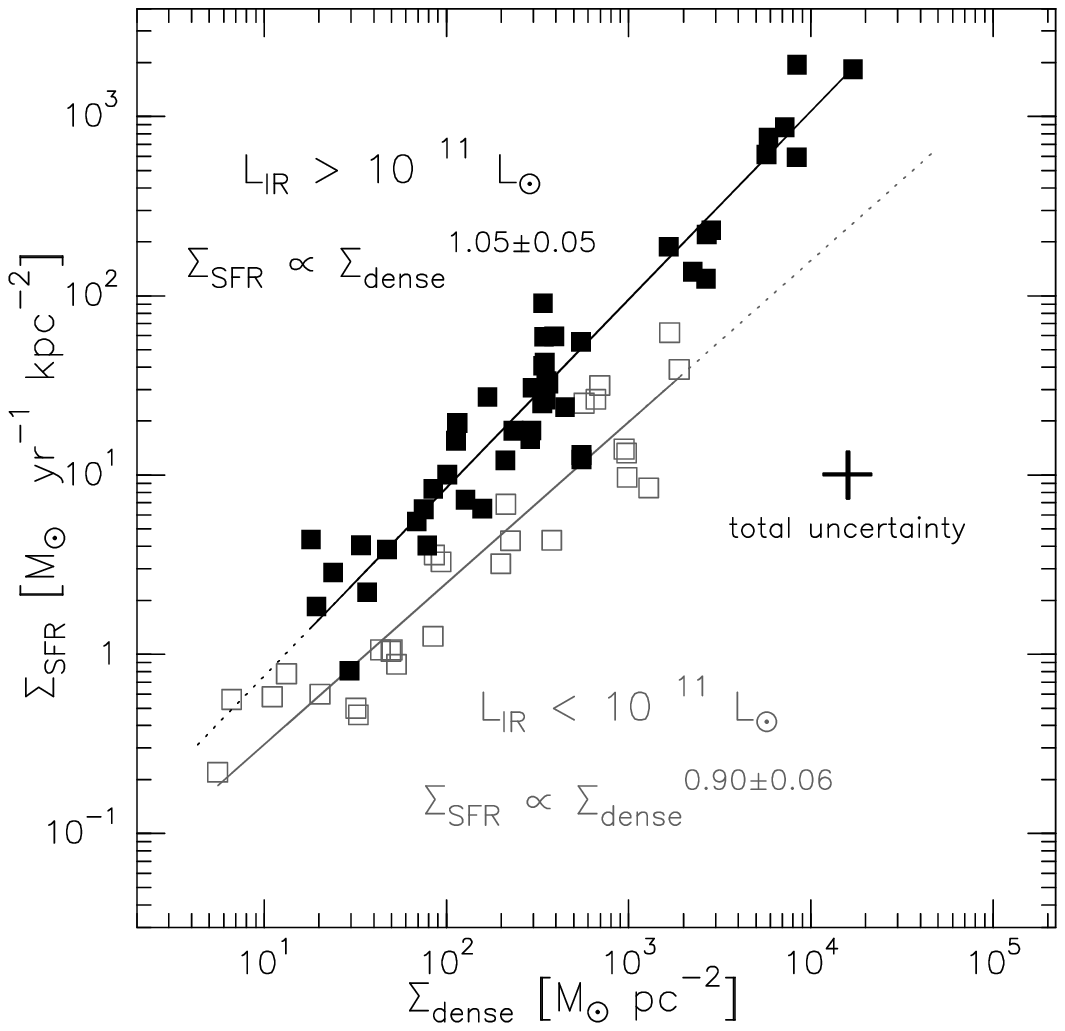}
         \includegraphics[width=6cm, angle=-90]{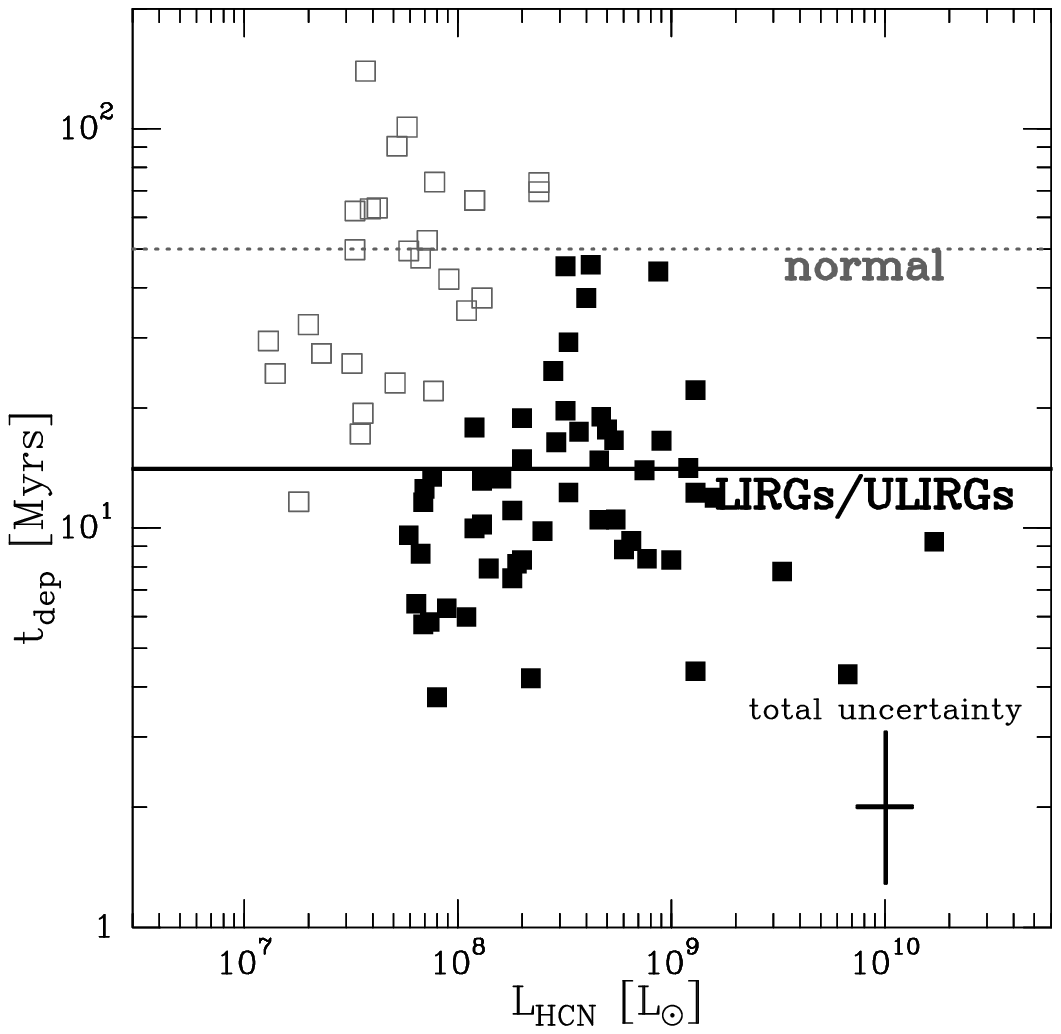}
   \caption{{\bf a)}~({\it left panel}) Star-formation rate surface density, $\Sigma_{\rm SFR}$, as a function of dense molecular gas surface  density, $\Sigma_{\rm dense}$, obtained with the {\em revised} conversion factors discussed in Sect.~2. The power law fits  to normal galaxies (gray line) and luminous infrared galaxies (black line) are shown.  Symbols as in Fig.~\ref{LFIR-LHCN}. {\bf b)}~({\it right panel}) We show the
   revised depletion time scale as a function of $L'_{\rm HCN}$  derived in normal and luminous infrared galaxies, as discussed in Sect. ~2. The horizontal lines 
 indicate the average value of the depletion time scale in normal galaxies (t$_{\rm dep}\sim$50$\pm$5~Myr) and LIRGs/ULIRGs (t$_{\rm dep}\sim$14$\pm$1.4~Myr).} 
              \label{KS-revised}
\end{figure*}

\begin{figure*}[th!]
   \centering
   \includegraphics[width=6cm, angle=-90]{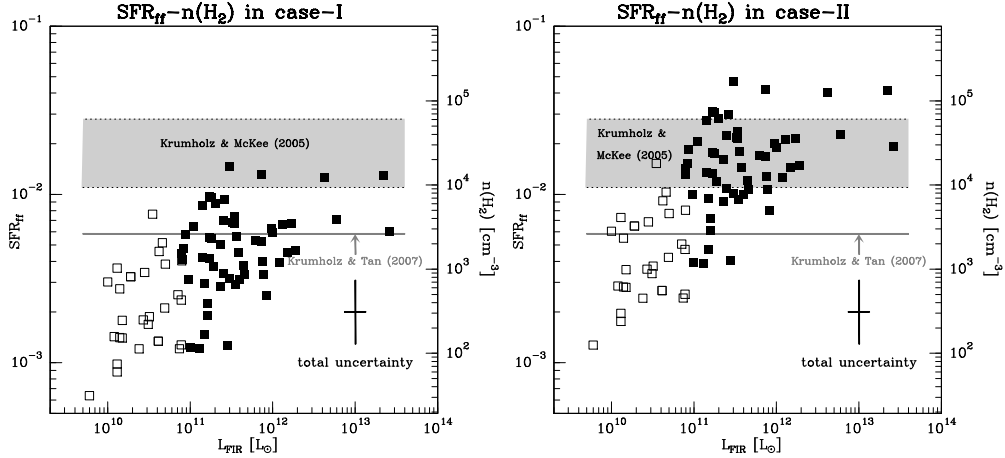}
   \caption{{\bf a)}~({\it left panel}) The SFR$_{\rm ff}$ (left Y axis) derived from the observed
    SFE$_{\rm dense}$ in different populations of galaxies assuming a gas density for HCN 
    n$_{\rm HCN}$(H$_2$)=3$\times$10$^4$cm$^{-3}$.  We show on the right Y axis the value of n$_{\rm HCN}$(H$_2$)
     derived from SFE$_{\rm dense}$ assuming a constant SFR$_{\rm ff}$=0.02, a value within the range determined by the star-formation model of \cite[Krumholz \& McKee~(2005)]{Kru05}. 
     We also highlight the average value of SFR$_{\rm ff}$$\sim$0.0058 determined by  \cite[Krumholz \& Tan~(2007)]{Kru07}. Symbols are as in Fig.~\ref{LFIR-LHCN}.   {\bf b)}~({\it right panel}) Same as {\bf a)} but obtained from the revised values of $\Sigma_{\rm dense}$  and $\Sigma_{\rm SFR}$ discussed in Sect.~6 of  \cite[Garc\'{\i}a-Burillo et al.~(2012)]{gb12}. } 
              \label{SFRff}
\end{figure*}

\section{Star formation laws of the dense gas: a new sample of LIRGs}

As illustrated in Fig.~\ref{LFIR-LHCN}a, with the addition of the new LIRG sample we confirm that the  $L_{\rm FIR}$-$L'_{\rm HCN(1-0)}$ correlation is superlinear (power law index $\sim$1.23$\pm$0.05). Furthermore we corroborate on a more solid statistical basis, compared to the work of  \cite{gc08}, that the average star-formation efficiency of the dense gas is a factor 
$\sim$2--3 higher in LIRGs/ULIRGs  ($\sim$1400$\pm$100$L_{\odot}\,$/(Kkms$^{-1}$pc$^2$)) compared to normal galaxies ($\sim$600$\pm$70$L_{\odot}\,$/(Kkms$^{-1}$pc$^2$)) (See Fig.~\ref{LFIR-LHCN}b).  If we use universal conversion factors for HCN and FIR luminosities, a 
fit to the full  sample yields a power index $n = 1.12 \pm 0.04$ for the implied Kennicutt-Schmidt  laws relating $\Sigma_{\rm SFR}$ and $\Sigma_{\rm dense}$.  However, LIRGs/ULIRGs and normal galaxies are not fully overlapping in this 
scatter plot (see Fig.~6 of \cite[Garc\'{\i}a-Burillo et al.~2012]{gb12}).  In order  to quantify if a dual fit applies to the power law, we split the sample into normal and IR luminous galaxies. A two-function power law fit  with indices close to unity qualifies as a 
much better description of the star-formation relation for the dense gas. This result is seen to be solid against the statistical 
biases inherent to this analysis (see discussion in Sect.~5 of \cite[Garc\'{\i}a-Burillo et al.~2012]{gb12}).

 The validity of a universal $M_{\rm dense}$ -- $L'_{\rm HCN(1-0)}$ conversion factor ($\alpha_{\rm HCN}$)  for normal galaxies and LIRGs/ULIRGs  has been questioned by a number of works, which favor lower $\alpha_{\rm HCN}$ in extreme starbursts compared to normal galaxies (e.g., \cite[Graci\'a-Carpio et al.~2008]{gc08}; see, however, \cite[Papadopoulos et al.~2012]{Pap12}).   If we account for the different conversion factors for HCN in extreme starbursts and for  the unobscured star-formation rate in normal galaxies the duality in star-formation laws is enhanced, as illustrated in Fig.~\ref{KS-revised}a. 
At the high end of $\Sigma_{\rm dense}$ values ($\sim$a few~10$^4$\,M$_{\odot}$pc$^{-2}$)  the normal galaxy law underpredicts $\Sigma_{\rm SFR}$ in IR luminous galaxies by about an order of magnitude. Furthermore, within the range of gas surface densities shared by normal galaxies and LIRGs/ULIRGs, $\Sigma_{\rm dense}$$\sim$2 10$^1$--2 10$^3$\,M$_{\odot}$pc$^{-2}$, the factor 3 to 5 disagreement between the two laws is a factor $\sim$3--4 times larger than the typical uncertainty on $\Sigma_{\rm SFR}$.
This result extends the more extreme bimodal behavior of star-formation 
laws, derived from CO molecular lines by two recent surveys (\cite[Daddi et al.~2010]{Dad10}, \cite[Genzel et al.~2010]{Gen10}), to the higher molecular densities probed by HCN lines.  On average, the revised depletion time scales for the dense molecular gas show a  significant difference  between LIRGs/ULIRGs  (t$_{\rm dep}\sim$14$\pm$1.4~Myr) and normal galaxies (t$_{\rm dep}\sim$50$\pm$5~Myr) (Fig.~\ref{KS-revised}b).

\section{Star-formation laws: observations versus models}

\cite[Garc\'{\i}a-Burillo et al.~(2012)]{gb12} have used the new HCN observations presented above to test a  key prediction of the star-formation model of \cite{Kru05}: the constancy of SFR$_{\rm ff}$, which is expected to be $\sim$0.02 up to the density regime explored by HCN observations. Figure~\ref{SFRff}a represents  SFR$_{\rm ff}$ as a function of L$_{\rm FIR}$ obtained from the SFE$_{\rm 
dense}$ values derived with universal conversion factors for L$_{\rm FIR}$ and L$_{\rm HCN}$.  To render a constant  SFR$_{\rm ff}$ compatible with the order of magnitude increase in SFE$_{\rm dense}$, the density of HCN clouds has to change notably from normal galaxies to LIRGs/ULIRGs.
Nevertheless, the values derived for  SFR$_{\rm ff}$ and/or   
n$_{\rm HCN}$(H$_2$) are well below the expected range both for normal galaxies and LIRGs/ULIRGs. 
The use of revised conversion factors alleviates the problem only in LIRGs/ULIRGs (Fig.~\ref{SFRff}b).
An overall satisfactory solution, which accounts for the observed trend in the L$_{\rm FIR}$/L'$_{\rm HCN}$ luminosity ratios, is only found provided that: 1)~SFR$_{\rm ff}$$\sim$0.0035 (i.e., a factor $\sim$6 lower than the canonical value of $\sim$2$\%$), 2)~n$_{\rm HCN}$(H$_2$) increases with L$_{\rm FIR}$ from $\sim$10$^4$cm$^{-3}$ in normal galaxies to $\sim$10$^6$cm$^{-3}$ in extreme starbursts, and 3)~$\alpha_{\rm HCN}$ is $\sim$2--4 times lower in LIRGs/ULIRGs. In summary: there is new observational evidence that the overall efficiency of the dense molecular gas in galaxies is significantly lower than initially postulated by \cite[Krumholz \& McKee~(2005)]{Kru05}. Furthermore, an agreement between observations and models requires a dramatic increase in the average gas densities from normal galaxies to extreme starbursts.

 Multi-line observations are a key to constrain the conversion factors of the different molecular tracers that are commonly used as  proxies for  $\Sigma_{\rm gas}$. These observations can also yield an estimate of volume gas densities. As recently argued by \cite[Krumholz et al.~(2012)]{Kru12} a simple, local, volumetric star-formation law with SFR$_{\rm ff}\sim$0.01 can dissolve the apparent bimodality found by \cite[Genzel et al.~(2010)]{Gen10} if different scale-heights  are assumed for the gas disks in normal galaxies and extreme starbursts. While this is a plausible explanation, the change in scale-heights  required to make bimodality disappear in star formation  laws remains  to be observationally validated.


\begin{thebibliography}{}

\bibitem[Alonso-Herrero et al.~(2006)]{Alo06} {Alonso-Herrero, A., Rieke, G.~H., Rieke, M.~J., et al.} 2006, \textit{ApJ}, 650, 835

\bibitem[Daddi et al.~(2010)]{Dad10} {Daddi, E., Elbaz, D., 
Walter, F., et al.} 2010, \textit{ApJ} (Letters), 714, L118 

\bibitem[Gao \& Solomon~(2004)]{Gao04} {Gao, Y., \& Solomon, P.~M.} 2004, \textit{ApJ}, 606, 271 

\bibitem[Garc{\'{\i}}a-Burillo et 
al.~(2012)]{gb12} {Garc{\'{\i}}a-Burillo, S., Usero, A., Alonso-Herrero, A., et al.} 2012, \textit{A\&A}, 539, A8 


\bibitem[Genzel et al.~(2010)]{Gen10} {Genzel, R., Tacconi, 
L.~J., Graci\'a-Carpio, J., et al.}  2010, \textit{MNRAS}, 407, 2091 

\bibitem[Graci{\'a}-Carpio et al.~(2008)]{gc08} {Graci{\'a}-Carpio, J., Garc{\'{\i}}a-Burillo, S., Planesas, P., et al.} 2008, \textit{A\&A}, 479, 703

\bibitem[Krumholz \& McKee~(2005)]{Kru05} {Krumholz, M.~R., \& McKee, C.~F.} 2005, \textit{ApJ}, 630, 250 

\bibitem[Krumholz \& Tan~(2007)]{Kru07} {Krumholz, M.~R., \& Tan, J.~C.} 2007, \textit{ApJ}, 654, 304 

\bibitem[Krumholz et al.~(2012)]{Kru12} {Krumholz, M.~R., 
Dekel, A., \& McKee, C.~F.}  2012, \textit{ApJ}, 745, 69 

\bibitem[Papadopoulos et al.~(2012)]{Pap12} {Papadopoulos, 
P.~P., van der Werf, P., Xilouris, E., Isaak, K.~G., 
\& Gao, Y.} 2012, \textit{ApJ}, 751, 10 

\bibitem[Schmidt~(1959)]{Sch59} {Schmidt, M.} 1959, \textit{ApJ}, 129, 243 






\end{thebibliography}
\end{document}